# High Photoluminescence Intensity of Heterostructure AlGaN-based DUV-LED through Uniform Carrier Distribution

Mohammad Amirul Hairol Aman[1], Faris Azim Ahmad Fajri[1], Ahmad Fakhrurrazi Ahmad Noorden[1,a], Mahdi Bahadoran[2], Suzairi Daud[3] and Muhammad Zamzuri Abdul Kadir[1]

[1] Centre for Advanced Optoelectronics Research (CAPTOR), Department of Physics, Kulliyyah of Science, International Islamic University Malaysia, 25200 Kuantan, Pahang, Malaysia

[2] Department of Physics, Shiraz University of Technology, Shiraz, Iran

[3] Laser Center, Ibnu Sina Institute for Scientific and Industrial Research, Universiti Teknologi Malaysia, Johor Bahru, Johor, Malaysia

[a] Author to whom correspondence should be addressed: fakhrurrazi@iium.edu.my

**Abstract:**

We report a numerical analysis of the variation of Aluminium (Al) composition in Al Gallium Nitride (AlGaN)-based Deep-Ultraviolet Light-Emitting Diode (DUV-LED) and its effects on the carrier concentration, radiative recombination, and photoluminescence (PL). Three different structures with different Al compositions are compared and analyzed. The radiative recombination of the DUV-LED is less efficient due to the imbalance of carrier distribution. The findings show that the uniform electrons and holes distribution significantly improve the radiative recombination for structure with a thin step-shaped quantum well (QW). The simulated structure emits a wavelength of 302.874 nm, categorized in the ultraviolet-B (UV-B) spectrum. Our results imply that carrier uniformity in QW is required to enhance the light intensity of DUV-LED. Remarkably, the uniformity enhances the PL intensity drastically, at least 300% higher than the other structures.

In recent years, the Deep-Ultraviolet Light-Emitting Diode (DUV-LED) has been under the limelight in the lighting and sensing industry, including the research field[1,2]. The DUV-LED has been very useful in various applications such as sterilization[3], indoor growing plant[4,5], disinfectant of virus[6], etc. There are a few factors affecting the performance of the DUV-LED such as the accumulation and distribution of carrier concentration in the active region[7], the design of the epitaxy layer[8] and the extra layer to facilitate the carrier transportation, such as the electron blocking layer (EBL)[9,10] and hole blocking layer (HBL)[10]. In general, the performance measurements of the LEDs are based on quantum efficiency and photoluminescence (PL); hence, these efforts focused on improving these qualities.

The PL is a process where the electron in higher energy states loses its energy and falls to a lower energy state. The energy conversion during the electron's transition is converted into emitting photons[11]. The PL spectrum of DUV-LED covers from 200 nm to 400 nm of light wavelength and can be categorized into three different ranges. UV-A covers from 320 nm to 400 nm, UV-B from 280 nm to 320 nm, and UV-C covers from 100 nm to 280 nm[12]. Countless efforts have been conducted to specify the desired wavelength and the PL intensity of DUV-LED. Most works focus on the types of material used to generate the UV spectrum, such as the near UVLED, Indium Gallium Nitride (InGaN)[13–15], and the DUV-LED Aluminium Gallium

Nitride (AlGaN)[16–21]. The width of the quantum well also plays a role in deciding the emitted UV-spectrum[22] More advanced research was performed to study the correlation between the PL with UV-spectrum and the thickness of QW as well as the quantum barrier (QB) by[23] Research on the composition of the material in the EBL and QB was conducted and focusing on the quantum efficiency of the LED instead of PL[23] The connection between the wavelength shift of the UV-spectrum in PL and the doping material also has been accomplished by[24] However, the incompatible of material composition is leading to low PL intensity.

The improvement of the DUV-LED active region has been attempted through various means to reduce electron leakage and increase the carrier concentration and radiative recombination. These improvements can be achieved by creating suitable carriers' confinement in the region. For instance, utilizing the graded-EBL and graded-quantum well[25], double-sided EBL[9], sandwiching the active region with p-type material[26], superlattice-EBL[27], step-graded multiple quantum barriers (MQB)[28,29], HBL[10], and combination of MQB-EBL[30]. Although the confinement of the carrier concentration is improved, the issue of radiative recombination in the active region continues to be an obstacle. The radiative recombination efficiency is low [31] even though the carriers are confined in the active region. As one of the motivational perspectives, this problem arises due to the non-uniform distribution in the active region among the quantum wells. In 2015, a successful method of achieving concentration uniformity was conducted using graded multiple quantum wells (GMQW) distinct from the conventional stepped QW (SQW)[32]. Nonetheless, the downside is that the GMQW shape is more complex to be fabricated than SQW, as the former requires a highly precise composition tuning. The inaccuracy of this composition tuning will damage the shape of a QW, thus deteriorating the carrier confinement efficiency lower than the SQW.

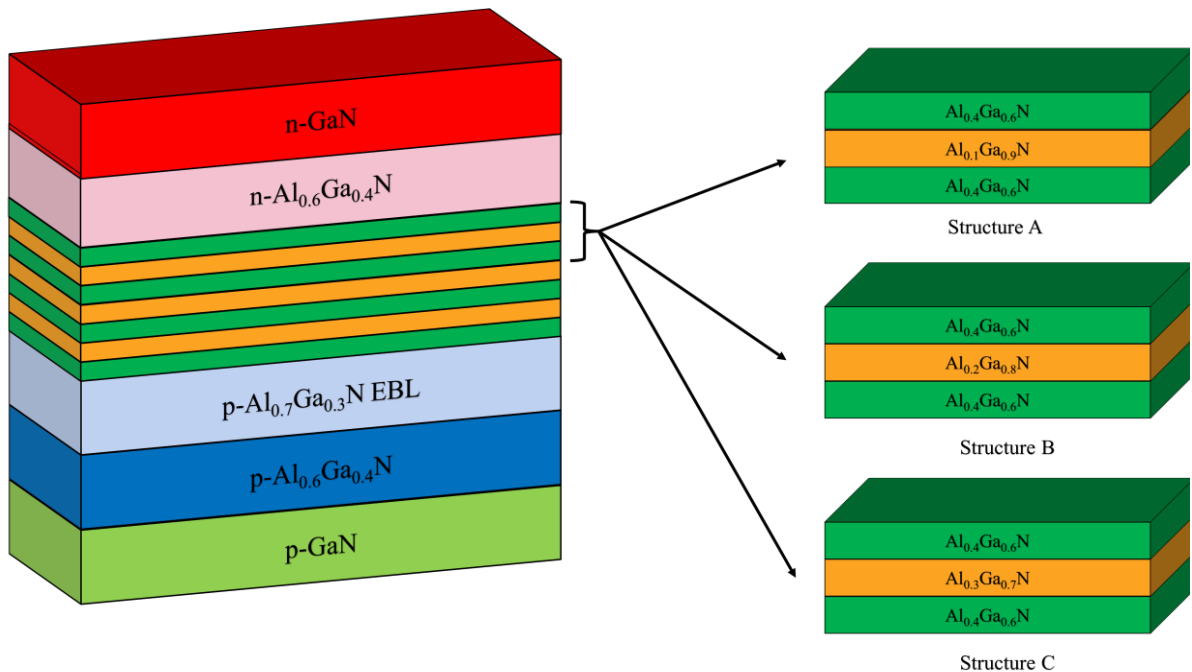

Figure 1: The epitaxy layer used in this simulation with the variation of the Aluminium composition in the quantum well.

In this work, utilizing the SQW, we analyze the composition of the Aluminium (Al) in the QW and Q and investigate the effects on the carrier concentration, radiative recombination, and PL to achieve uniform carrier distribution. The Al-composition in QW varies with a constant QB Al composition in the AlGaN-based DUV-LED chip. The simulated epi-layer is GaN/AlGaN material, and it has been performed using 1D-DDCC software[33] developed by Yuh-Renn Wu. Three different structures of $Al_xGa_{1-x}N$-based LED have been chosen as the Al composition of the QW varied from 0.1 to 0.3, shown in Figure 1. The epitaxy layer of the DUV-LED consists of n-GaN with 100 nm thickness followed by n-$Al_{0.6}Ga_{0.4}N$ with 150 nm thick. Both layers were doped with $5 \times 10^{18}$ cm$^{-3}$ Silicon. Then, an active region consists of three AlGaN QWs and four AlGaN QBs. The thickness of the QWs and QBs were set to 4 and 10 nm, respectively. The Al composition was varied according to Table 1. The next layer is a 20 nm p-$Al_{0.7}Ga_{0.3}N$ electron blocking layer with $3 \times 10^{19}$ cm$^{-3}$ Magnesium (Mg) doping, 150 nm p-$Al_{0.6}Ga_{0.4}N$ with $3 \times 10^{19}$ cm$^{-3}$ Mg-doped and lastly, 100 nm p-GaN with $1 \times 10^{20}$ cm$^{-3}$ Mg doping. The temperature was set to 300K, with 8V of applied voltage and the other parameters of the system can be referred to in[34–37].

In 1D-DDCC, the Drift-Diffusion Model (DDM) accurately approximates the energy band levels. The DDM comprises the semi-classical model, which describes the electron transport in semiconductors through Boltzmann's transport equation, and the current-continuity equations[38,39] The current-continuity equations and Schrodinger-Poisson equation (a model for carriers' wave function and distribution) are as the following

$$\vec{J_c} = qc(x)\mu_c\vec{E_x} + qD_c\frac{dc}{dx} \qquad (1)$$

$$\left\{-\frac{h^2}{2}\left(\nabla \cdot \frac{\nabla}{m^*}\right) + U\right\}\psi(x) = E\psi(x) \qquad (2)$$

Where $\vec{J_c}$ is the current density, $c$ is the carrier concentration, $\mu_c$ is the carrier mobility, and $U$ is the potential energy supplied to the system as the input for the simulation. The complete formulations[39] are linearized to 1-dimensional as they are solved self-consistently with uniform nodes to obtain the solution of carrier concentration and fermi level in spatial distribution. These solutions contribute to calculating energy bands, carrier recombination rates, and PL spectrum.

Table 1: The composition of the Aluminium for QW was varied.

| Parameter | Structure A | Structure B | Structure C |
|---|---|---|---|
| Quantum Well | 0.1 | 0.2 | 0.3 |
| Quantum Barrier | 0.4 | 0.4 | 0.4 |
| Difference between QW and QB, $\Delta_{comp}$ | 0.3 | 0.2 | 0.1 |

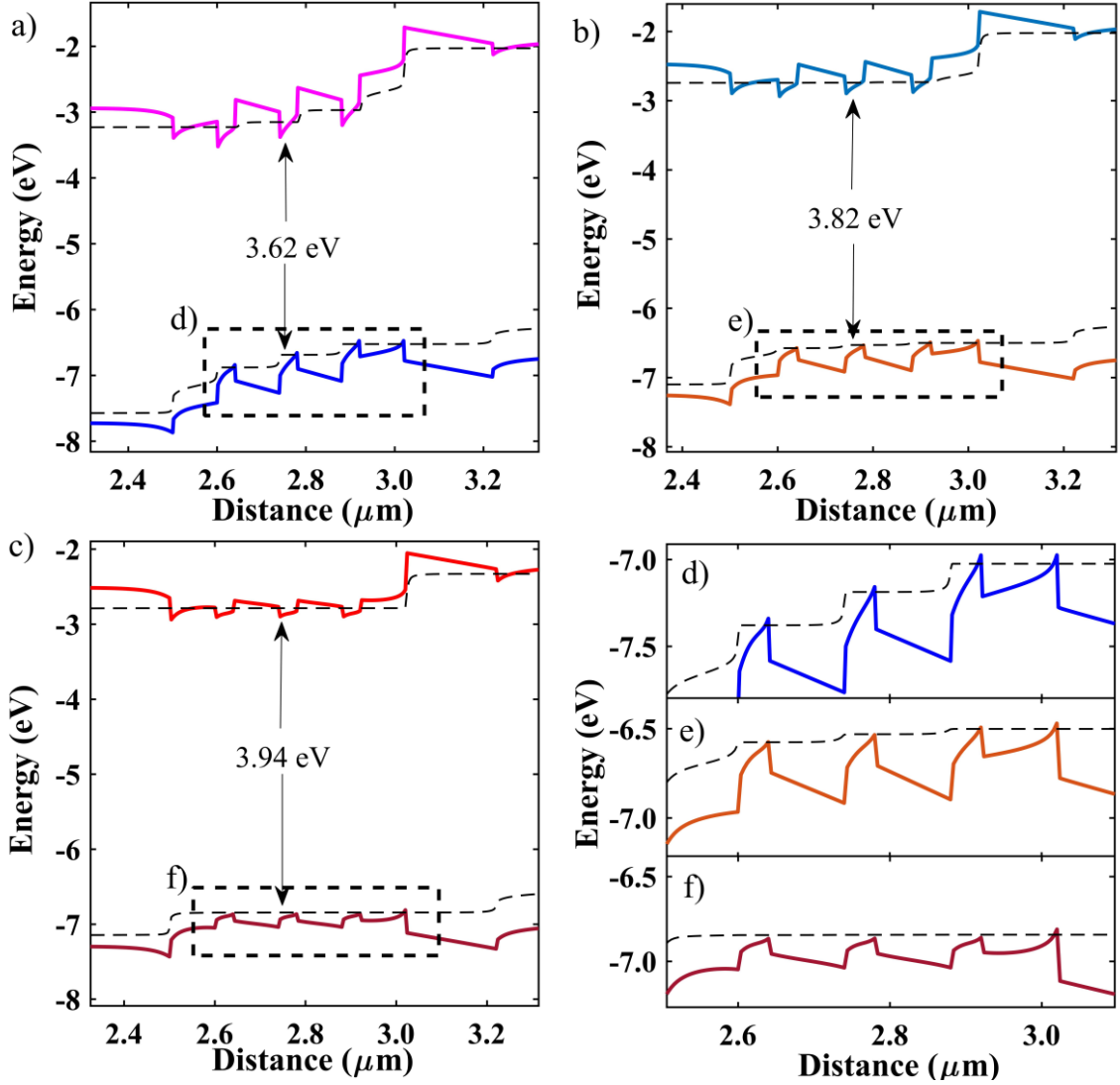

Figure 2. The energy band diagram of structures a) A (0.3 $\Delta_{comp}$), b) B (0.2 $\Delta_{comp}$), and c) C (0.1 $\Delta_{comp}$). The dashed line in the band diagram represents the quasi-fermi levels, and the dashed boxes are the magnified active regions for each structure.

Figure 2 depicts the band diagrams for all structures. The findings show that as the composition of the Al increases, the depth of the QW slightly decreases. As the difference between the QW and QB heights decreases, the holes' quasi-fermi level (dashed line) straightens across with the quantum well, indicating that the probability of the carrier presence at that space is uniform. This is due to the carriers' disorder effects, i.e., the second-order perturbation in the model of virtual crystal approximation[40]. This behaviour strongly agrees as it also affects the bandgap energy, $E_g$, which is a simplified quadratic form approximation based on the bowing parameter $C$ [34]

$$E_g(A_{1-x}B_x) = (1-x)E_g(A) + xE_g(B) - x(1-x)C \qquad (4)$$

The average energy gap of structures A, B, and C are 3.62 eV, 3.82 eV, and 3.94 eV, respectively. Therefore, altering the composition of Al can directly generate the desired energy band gap. This can be understood by considering AlGaN as the ternary alloy and the bandgap energy calculated in a simplified quadratic equation.

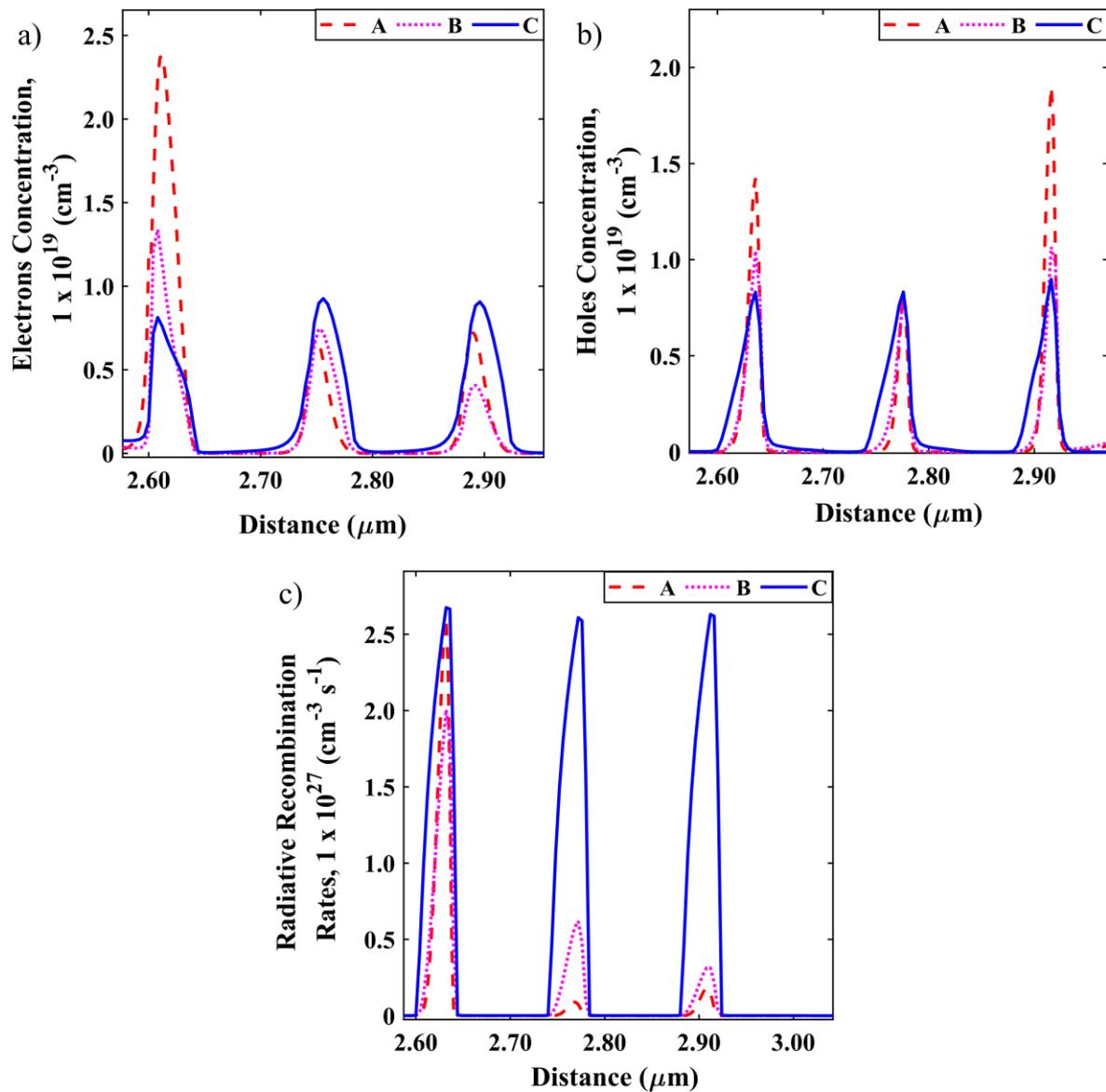

Figure 3: The a) electron concentration, b) hole concentration, and c) radiative recombination rates in the active region of all structures

Generally, it considers the bowing parameter for the alloys that involve the effects originating from the presence of different carriers, i.e., anions and cations. It has been calibrated accordingly to fit the experimental data[41], and the mathematical model has been developed for the virtual-crystal approximation based on the electronegativity theory, where the essence of electronegativity theory conveys the tendency of atoms to attract electrons[42] This leads to the alteration of the energy band level. Furthermore, the quasi-fermi level close to the altered band level can provide a high probability density of carrier[39]. This can be implied by the idea deduced from Figure 3.

Figure 3a) shows the electron concentration in the active region for structures A, B, and C. There is a noticeable electrons concentration for Structure A in the first QW positioned from 2.60 to 2.65 um. The electrons concentration decreases significantly in the third QW, followed by the second QW. As the Al-composition changes in Structure B, the pattern of the carrier distribution changes. Next, the electrons concentration gradually decreases from the first to the third QW. There is a considerable distinction in electrons concentration for structure C compared to structures A and B. The electrons distributions in all three QWs are close to uniformity. These can be inferred from the pattern of the quasi-fermi level in the band diagram in Figure 2, where the quasi-fermi level can be interpreted as the probability of the carrier having the specific energy level[39].

Furthermore, based on Figures 2 d), e), and f), the quasi-fermi level in the band diagram for structure A shows a stair-shaped pattern as well as in structure B. However, the stair shape for the quasi-fermi level of the active region in structure B (refer to Figure 2e) is relatively less than in structure A. This is the reason why the electrons concentration in Structure B gradually decreases. For structure C, there is no 'staircase' pattern in the active region, indicating that the electrons concentration for structure C is evenly distributed among the QWs. The 'shallow' QW in structure C is the main factor of the uniform carrier distribution. Contrarily, in structures A and B, more energy is required to pass through the QB. Hence, the carrier accumulation is the highest in the first QW.

Figure 3b) shows the holes concentration in the active region. The distribution also has a similar scenario as the electrons concentration, but the situation occurs starting at the opposite side. For structure A, the highest holes accumulation is in the third QW, followed by the first QW and the second QW. In structure B, the holes concentration in the three QWs has the same pattern as structure A but is slightly lower in magnitude. On the other hand, the holes concentration for Structure C is distributed uniformly, equivalent to the electron distribution in the three QWs.

Figure 3c) exhibits the structures' radiative recombination rate in the active region. Radiative recombination refers to the recombination of electron-hole pairs where the energy loss from the recombination will be released in photon emission[43,44] The non-radiative recombination is also the recombination of electron-hole. Still, instead of emitting photons, it generates unfavorable phonons since it will reduce the efficiency [45,46]. The radiative recombination rates for structures A and B are not uniform. The recombination occurrence for structure A in the second QW is the lowest, while for structure B, the third QW is the lowest. As previously mentioned, these phenomena can be traced back to the magnitude of electrons and holes in the QWs.

The recombination rates for structures A and B in the QWs also differ significantly. The radiative recombination in Structure C is uniform in all three QWs due to the uniform electrons and holes distribution in the three QWs. Although the carrier concentration in the QW of structure C is lower than in structures A and B, since the carrier distribution is uniform among the three QWs, it increases the radiative recombination rate. A DUV-LED with high radiative recombination is sought after[25,47,48], and thus, among the three structures simulated, structure C offers the highest radiative recombination in all three QWs.

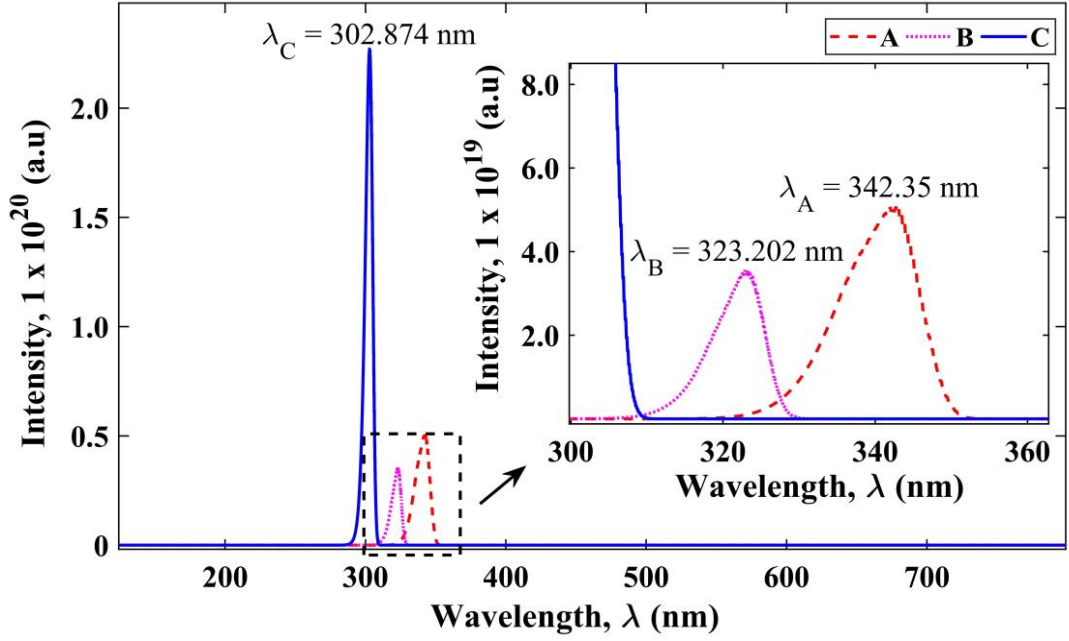

Figure 4: The photoluminescence of structures A, B, and C.

Figure 4 shows the PL of structures A, B, and C. All the wavelengths generated is within the UV spectrum but has different light intensity. The intensity of the PL for structure C is at least four times higher than structure A and B. PL with high intensity also implies that the structure has high radiative recombination of electron-hole pair[49] PL intensity is defined[50] by the product of distribution of carriers (based on Boltzmann distribution function) and the joint density of states as

$$I(\lambda) = \left[\frac{1}{2\pi^2}\left(\frac{2m_T^*}{h^2}\right)^{3/2}\sqrt{E(\lambda) - E_g^*}\right]\exp\left(-\frac{E(\lambda)}{kT}\right) \quad (3)$$

$m_T^*$ is the combined effective mass of the carriers. The light intensity is calculated based on the effective bandgap energy, $E_g^*$. The effective photon energy is reduced due to several compensating factors, including the activation energy, phonons generation, and thermal ionization energy. Moreover, according to [36,51], the compensation for PL emission involves the monotonic decrease of emission intensity resulting from the dopants' thermal dissociation. From the classification of the UV spectrum mentioned previously, the light emitted by structures A, B, and C can be categorized in Table 2. The UV-A wavelengths can be applied to agriculture as indoor plant growth [4,5], and in the medical field such as phototherapy[2] and radiation curing via radical polymerization[52]. The UV-B wavelength is also helpful in agriculture, such as becoming a eustressor to increase the photosynthesizing of the plant [1,4].

Table 2: The difference in Aluminium composition between quantum well and barrier, bandgap energy, the wavelength emitted, and type of UV generated for structures A, B, and C.

| Structure | Difference of Al-Composition in QW/QB, $\Delta_{comp}$ | Bandgap Energy (eV) | Peak Wavelength (nm) | Type of UV |
|---|---|---|---|---|
| A | 0.3 | 3.62 | 342.350 | UV-A |
| B | 0.2 | 3.82 | 323.202 | UV-A |
| C | 0.1 | 3.94 | 302.874 | UV-B |

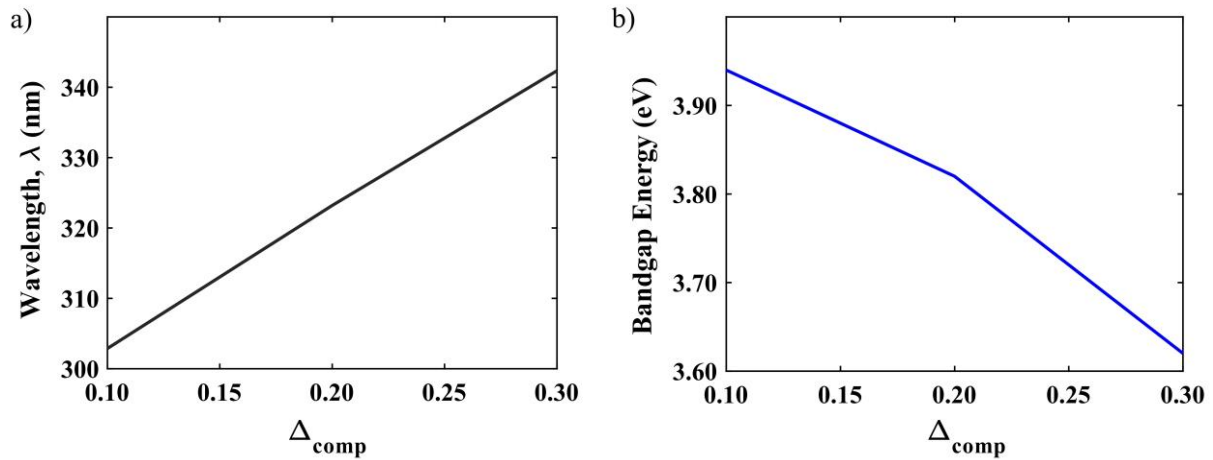

Figure 5a) The peak wavelength of the emitted light, and b) The bandgap energy varies with $\Delta_{comp}$ which is the Al composition difference between QWs and QBs.

Figure 5a) shows that as the difference between QW and QB increases the wavelength of PL emitted by the structure increases. This is because the Al composition difference between QWs and QBs affects the conduction and valence band, where the bandgap energy is also affected. It can be observed in Figure 5b) that when the Al composition difference between QWs and QBs increases, the energy bandgap also increases, which is why the wavelength of the PL is decreased. With this analysis, the wavelength of the PL can be adjusted by modifying the Al composition. Besides, as the radiative recombination in the quantum wells increases, the high intensity of PL can be achieved[49]; hence, in the context of PL, having a structure with uniform radiative recombination as Structure C is the best selection for high-intensity DUV-LED applications.

In conclusion, the three GaN/AlGaN structures with different Al-composition of the non-complex SQW were analyzed accordingly for high-intensity PL of DUV-LED. Structures A, B, and C each have advantages, as the PL wavelengths are 342.35 nm, 323.202 nm, and 302.874 nm, respectively. The uniform carrier distribution is obtained by decreasing the difference between QW and QB, i.e., structure C. Using this structure, the radiative recombination has been improved significantly by four times of intensity compared to structures A and B. The high intensity of PL also indicates high radiative recombination, reflecting a highly efficient structure for DUV-LED. Therefore, the high-intensity DUV-LED applications have the potential to become the cornerstone of many fields and disciplines.